\documentclass[journal]{IEEEtran}
\IEEEoverridecommandlockouts 

\usepackage{cite}
\usepackage{siunitx}
\usepackage{float}
\usepackage{amssymb}
\usepackage[table,xcdraw]{xcolor}

\usepackage{algpseudocode}
\usepackage{algorithm}

\usepackage{amsthm}
\usepackage{flushend}

\newtheorem{mydef}{Definition}

\ifCLASSINFOpdf
   \usepackage[pdftex]{graphicx}
   \graphicspath{{./Images/}}
   \DeclareGraphicsExtensions{.pdf,.jpeg,.png}
\else

\fi

\usepackage[cmex10]{amsmath}
\usepackage[hidelinks]{hyperref}
\usepackage{multirow}
\usepackage{booktabs,colortbl}
\usepackage[format=hang]{caption}

\usepackage[version=3]{mhchem} 

\makeatletter
\newcommand*{\rom}[1]{\expandafter\@slowromancap\romannumeral #1@}
\makeatother

\begin{document}
\title{Fast Stability Scanning for Future Grid Scenario Analysis} 

\author{Ruidong Liu,~\IEEEmembership{Student Member IEEE}, Gregor~Verbi\v{c},~\IEEEmembership{Senior~MIEEE}, Jin Ma,~\IEEEmembership{Member~IEEE}
\thanks{Ruidong Liu, Gregor~Verbi\v{c}, and Jin Ma are with the School of Electrical and Information Engineering, The University of Sydney, NSW 2006, Australia, (e-mails: ruidong.liu@sydney.edu.au, gregor.verbic@sydney.edu.au, jin.ma@sydney.edu.au)}
}

\maketitle

\begin{abstract}
Future grid scenario analysis requires a major departure from conventional power system planning, where only a handful of most critical conditions is typically analyzed. To capture the inter-seasonal variations in renewable generation of a future grid scenario necessitates the use of computationally intensive time-series analysis. In this paper, we propose a planning framework for fast stability scanning of future grid scenarios using a novel feature selection algorithm and a novel self-adaptive PSO-k-means clustering algorithm. To achieve the computational speed-up, the stability analysis is performed only on small number of representative cluster centroids instead of on the full set of operating conditions. As a case study, we perform small-signal stability and steady-state voltage stability scanning of a simplified model of the Australian National Electricity Market with significant penetration of renewable generation. The simulation results show the effectiveness of the proposed approach. Compared to an exhaustive time series scanning, the proposed framework reduced the computational burden up to ten times, with an acceptable level of accuracy.
\end{abstract}

\begin{IEEEkeywords}
Future grids, scenario analysis, stability scanning, time-series analysis, small-signal stability, voltage stability, machine learning, clustering, feature selection.
\end{IEEEkeywords}

\vspace{-0.15cm}
\section{Introduction}

Power systems are undergoing a major transformation driven by the increasing uptake of renewable energy sources, DC power transmission, and the decentralization of electric power supply underpinned by the information and communication technologies and demand-side technologies, like rooftop PV, energy storage, home energy management systems, and electric vehicles. How future grids will look like, however, is still uncertain as the evolution depends not only on technological development but also on the regulatory environment. 
Therefore, one of the challenges associated with future grid planning is that the structure of a future grid cannot be simply extrapolated from the existing one.
As an example, the emergence of prosumers\footnote{Consumer with generation (e.g.rooftop-PV) and battery storage (\textbf{pro}ducer-con\textbf{sumer}).} might  change the demand profile, which results in a significantly different stability performance, as demonstrated in \cite{Marzooghi2016a}. Instead, for future grids planning, several possible evolution paths need to be accounted for. Future grid planning thus requires a major departure from conventional power system planning, where only a handful of the most critical scenarios is analyzed. To account for a wide range of possible future evolutions, scenario analysis has been proposed in other industries, e.g. in finance and economics \cite{LearningFromTheFuture_1998}, and in energy \cite{Foster2013}. As opposed to the conventional power system planning, where the aim is to find an optimal transmission and/or generation expansion plan for an existing grid, the aim in scenario analysis is to analyze possible evolution pathways to inform power system planning and policy making. Given the uncertainty associated with long-term projections, the focus of future grid scenario analysis should focus on analyzing what is technically possible, although it might also consider an explicit costing \cite{Elliston2016}. Therefore, future grids' planning may involve large amount of scenarios and the existing planning tools may no longer suitable.

Future grid analysis is a growing research area. Melbourne Energy Institute \cite{PHearpsMWright2010} have proposed a possible plan for a future Australian grid relying 100\% on renewable energy sources (RES). The Centre for Energy and Environmental Markets at the UNSW \cite{Elliston2012, Elliston2013} has shown for the Australian National Electricity Market (NEM) that balancing a 100\% RES power system is technically possible. The PJM study \cite{Budischak2013} has shown that the PJM network can be powered 90-99.9\% of the time entirely on RESs, at a cost comparable to today's. The existing studies, however, only focus on balancing and use a simplified copper plate model of the transmission network. They also neglect stability analysis, which limits their value.

Stability analysis is an important task in power system planning. In conventional stability analysis, only a small number of worst-case critical conditions is typically analyzed. If stable under those conditions, the system is assumed stable in all possible credible operating conditions. The selection of the critical conditions is most often based on the historical performance, and planners' experience and judgment \cite{Bebic2008, Quintero2015, Eftekharnejad2013, Knuppel2012, Klein1991}. In power systems with significant penetration of intermittent RES, the generation dispatch and the associated power flows change many times throughout the day and often follow rather different seasonal patterns, which renders past operational experience of limited value. Although the authors of a future grid study \cite{Miller2015} selected a few critical operation points for stability analysis, they also pointed out that there is no guarantee that these cases are necessarily the most difficult ones.
Chronological time series scanning offers a way for the stability analysis of a power system with a constantly varying operating conditions, and to capture the inter-seasonal variations in renewable generation. With time series scanning, it is possible to capture stability performance over a long horizon. The authors in \cite{Vittal2010} have demonstrated the value of using time-series analysis for steady-state voltage stability analysis of a power system with high penetrations of wind. They have shown that in contrast to traditional power systems without intermittent generation, in a system with a high RES penetration, the worst case operating point shifts. The time-consuming time-series simulation, however, was not discussed in \cite{Vittal2010}. Instead, the worst case points were manually picked  from several years worth of data, and the simulations were performed around these points to reduce the computational burden.

To the best of our knowledge, the Future Grid Research Program funded by the Australian Commonwealth Scientific and Industrial Research Organisation (CSIRO), is the first to propose a comprehensive modeling framework for future grid scenario analysis. The aim of the project is to explore possible future pathways for the evolution of the Australian grid out to 2050 by looking beyond simple balancing. 
To this end, a simulation platform has been proposed in \cite{Marzooghi2014} that consists of a market model, power flow analysis, and stability analysis. Preliminary results have shown, however, that time-series scanning over a one-year horizon is computationally very expensive. To speed-up the computation, we propose a machine learning (ML) based framework for fast stability scanning. 
The efficacy of the framework is demonstrated on a simplified 14-generator model of the Australian National Electricity Market.

The contribution of the paper is twofold: (i) we propose a planning framework for fast stability scanning of future grid scenarios, which makes it possible to analyze a large number of scenarios with a moderate computational effort; 
(ii) a novel self-adaptive PSO k-means clustering algorithm that considers both the adjusted feature ranks and wights for clustering and optimal selection of the number of clusters.

The rest of the paper is organized as follows: Section II outlines the simulation platform for future grid scenario analysis. Section III gives an overview of the application of ML in power systems and describes the pertinent ML algorithms. Section IV proposes a novel fast stability scanning framework. In Section V, the efficacy of the proposed framework is demonstrated on a simplified 14-generator network model of the Australian National Electricity Market. Section VI concludes the paper.

\vspace{-0.15cm}
\section{Simulation Platform}
We use the simulation platform for future grid scenario analysis originally proposed in \cite{Marzooghi2014} as the basis, summarized in Algorithm \ref{platform}. The platform consists of four modules: (i) scenario generation, (ii) market simulation, (iii) load flow analysis, and (iv) stability analysis, described in more detail later. The other three modules remain the same.

\begin{algorithm}
\caption{Future grid scenario analysis.} \label{platform}
\textbf{Input}: Network data, generation data, wind, solar and demand traces for each scenario $s \in \mathcal{S}$ in the studied year.\\
\textbf{Output}: Stability indices for each time slot $t \in \mathcal{T}$, for each scenario $s \in \mathcal{S}$.
\begin{algorithmic}[1]
\For{$s \gets 1, \left| \mathcal{S} \right|$}
	\For{$t \gets 1, \left| \mathcal{T} \right|$}
		\State{Market simulation (generation dispatch);}
		\State{Load-flow analysis;}
	\EndFor
	\For{$t \gets 1, \left| \mathcal{T} \right|$}
		\State{Stability analysis (voltage, angle, frequency);}
	\EndFor
\EndFor
\end{algorithmic}
\end{algorithm}

\vspace{-0.15cm}
\subsection{Test System}
We use a modified 14-generator IEEE test system that was initially proposed in \cite{Gibbard2010} as a test bed for small-signal analysis. The system is loosely based on the Australian National Electricity Market (NEM), the interconnection on the Australian eastern seaboard. The network is stringy, with large transmission distances and loads concentrated in a few load centers. It consists of 59 buses, 28 loads and 14 generators, each representing a power station consisting of between 2 to 12 units, resulting in a total of 74 synchronous machines. The single-line diagram of the test-bed is illustrated in Fig. \ref{fig:1}, in which Areas 1 to 5 represent Snowy Hydro (SH), New South Wales (NSW), Victoria (VIC), Queensland (QLD) and South Australia (SA), respectively. Areas 1 and 2 are electrically closely coupled, hence the system has four distinct areas.

\vspace{-0.15cm}
\subsection{Scenario Description (Line 1 in Algorithm \ref{platform})}
\begin{figure}[h]
\centerline{\includegraphics[height=13.8cm]{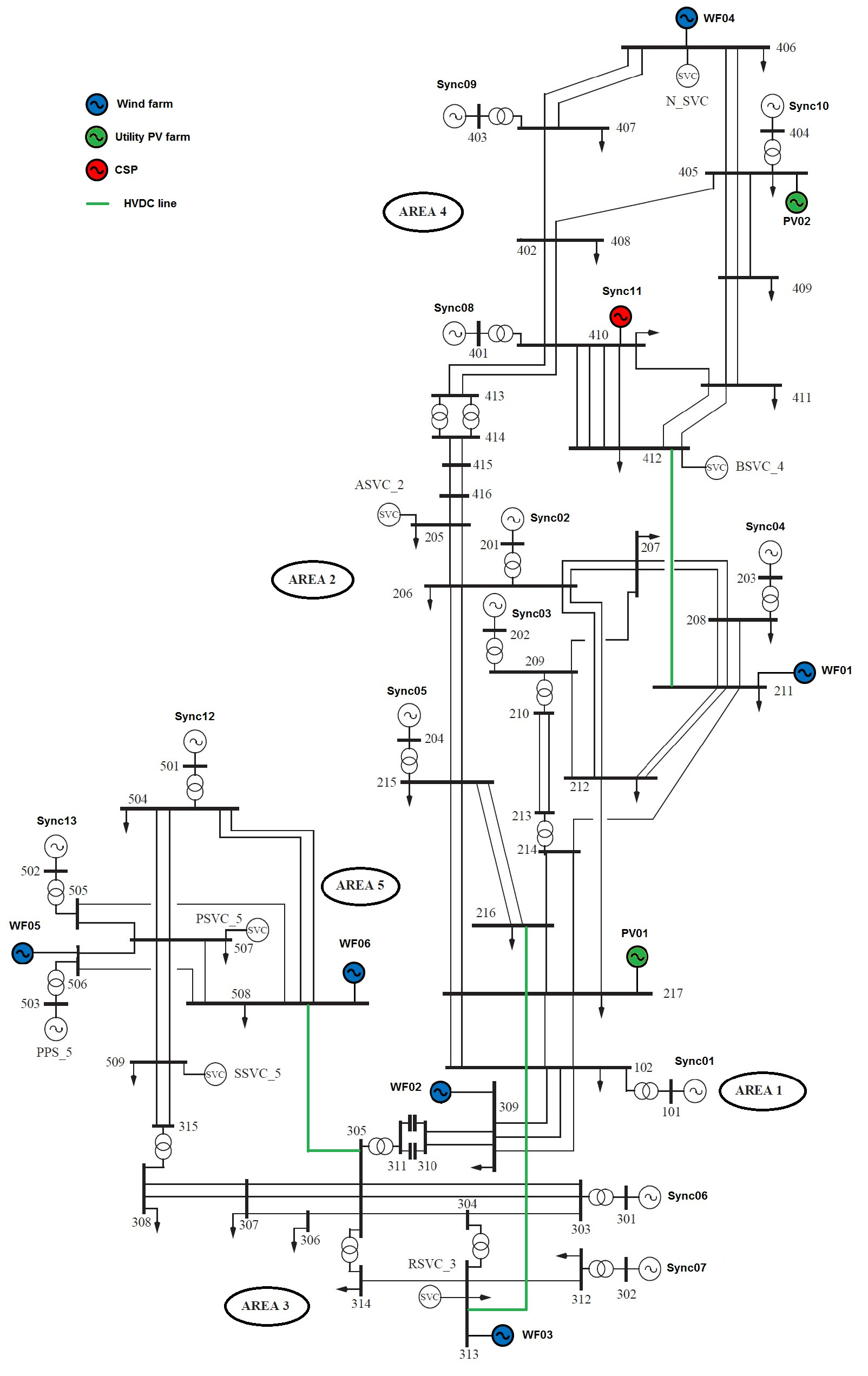}}
\caption{14-generator test system.}
\vspace{-0.6cm}
\label{fig:1}
\end{figure}

Given that the focus of the paper is fast stability scanning, we only analyze one future grid scenario. 
We augmented the test system by replacing conventional synchronous generators at selected buses with wind farms (WF) and PV farms, and a concentrated solar thermal plant (CSP), as shown in Fig. \ref{fig:1}, resulting in 30\% RES energy penetration. To increase the transfer capacity of the network, we added HVDC links between buses 412 and 211, 216 and 313, 305 and 508, reinforced the existing AC transmission corridors and added static var compensator to improve voltage control.
We used wind, solar and demand predictions for the year 2030 from the Australian Energy Market Operator's National Transmission Network Development Plan \cite{AEMO2012}.

\vspace{-0.15cm}
\subsection{Time-series Analysis (Lines 2-5 in Algorithm \ref{platform})}
Time-series analysis consists of market simulation and load-flow analysis using the generation dispatch results. To capture the inter-seasonal variations in renewable generation and the demand, we need to analyze a full year, which results in $\left| \mathcal{T} \right| = 8760$ assuming hourly resolution.

\subsubsection{Market Model (Line 3 in Algorithm \ref{platform})}
The aim of the market model is to emulate the outcome of an efficient electricity market without assuming any particular market structure. The model is based on a unit commitment problem aiming to minimize total electricity generation cost, and is subject to the following constraints: power balance, spinning reserve, power generation limit, start-up and shut-down constraints, ramp rate limits, generator minimum up time restrictions, and generator minimum down time restrictions. To achieve an acceptable computational performance, the resulting mixed-integer optimization problem is solved using a rolling horizon approach with hourly resolution. The decision horizon is two days, where the solution for the first day is retained, and the solution of the next day overlaps with the next two-day horizon. We assume that generators bid at their respective short-run marginal cost, which we assume to be zero for RES. A more complete description of the model is given in \cite{Riaz2016}.

\subsubsection{Load-flow Analysis (Line 4 in Algorithm \ref{platform})}
Load-flow analysis uses the dispatch results of market simulation and the load traces from \cite{AEMO2012}. RES are assumed to operate in a voltage-control mode. With hourly resolution, we obtain 8760 operating points, or instances, representing the year 2030. Each operating point is represented by a set of steady-state power system variables, or features. The operating points resulting from the time-series analysis are used for stability analysis.

\vspace{-0.15cm}
\subsection{Stability Analysis (Lines 6-8 in Algorithm \ref{platform})}
In this paper, we focus on small-signal and static voltage stability, although the simulation platform can also cover large-disturbance angle (transient) stability and frequency stability \cite{Shabir2016}.

\subsubsection{Modal Analysis}
Small disturbance (or small-signal) rotor angle stability is concerned with the ability of a power system to maintain synchronism under small disturbances \cite{Kundur2004}. Small signal stability problems are usually due to lack of damping. Inter-area oscillation modes may cause large power swing across inter-connectors and can lead to system collapse or splitting. In this study, the system exhibits a poorly damped inter-area mode between NSW and QLD, which is the focus in stability scanning. We use modal analysis of a power system model linearized around the current operating point.

\subsubsection{Steady-state Voltage Stability}
Voltage stability refers to the ability of a power system to maintain steady voltages at all buses in the system from a given initial operating condition\cite{Kundur2004}. Voltage stability problems are typically associated with lack of reactive power support, which can result from heavily load transmission lines. In systems with high RES penetration, as in this study, this is of particular importance given the constantly varying power infeed.
Several stability indices have been proposed for voltage stability assessment, giving a measure of the distance of the current operating point from the voltage collapse point \cite{VanCutsemVournas_1998}. 
In this study, we used the loadability margin for stability scanning. 

\vspace{-0.15cm}
\section{Clustering and Feature Selection}
Before describing the proposed framework for fast stability scanning, we first give an overview of the application of ML in power systems and describe the two pertinent ML algorithms, i.e. k-means for clustering and ReliefF for feature selection.

The application of machine learning (ML) for stability analysis has attracted a significant attention in recent years \cite{Zhou2010, Shayanfar2012, Xu2012, Amjady2007}. In online dynamic security assessment (DSA)\footnote{Online security assessment involves both dynamic and static security assessment. The term dynamic security assessment is usually used to denote both.}, ML is used for classification, to map a system operating condition into a suitable stability index, for example for voltage stability \cite{Zhou2010, Shayanfar2012} and on-line transient stability assessment \cite{Xu2012, Amjady2007}.  
The classification of a system security status consists of three steps: (i) a large database is generated using time-domain simulation to create a training set; (ii) a set of features that best describe an operating condition is selected as the inputs of the classifier; and (iii) the classifier is trained using an appropriate tool, e.g. an artificial neural network \cite{Zhou2010, Shayanfar2012, Aboytes1996}, a support vector machine \cite{Mohammadi2009}, or a decision tree \cite{He2013a}.
To cover a large amount of possible operation conditions and to achieve an acceptable level of accuracy, however, the training set is normally very big---thousands of operation points for a relatively large conventional systems with no RES \cite{Xu2012a}. 
One possible way to address the problem is to reduce the size of the training data set by limiting or fixing the load or generation variation range for imminent hours only \cite{Aboytes1996,He2013a}. 
For a study of a future power system with high RES penetration, the possible operating space is much larger than a conventional one. Therefore, a direct application of the existing ML algorithms becomes infeasible. Instead, as proposed in this paper, clustering is required to reduce the number of operating points for stability scanning.

\vspace{-0.15cm}
\subsection{Clustering}
Clustering is the task of grouping a set of objects into clusters based on their similarity \cite{Xu2005}. A cluster is described by its internal homogeneity and the external separation, i.e., patterns in the same cluster should be similar to each other while patterns in different clusters should not \cite{Xu2005}. When clustering a large amount of data, their similarity is usually expressed as a distance. After clustering, all elements within a particular cluster can be represented by the center of this cluster or a cluster centroid. In power systems, clustering is a popular ML algorithm used for dimensionality reduction. It has been used in load forecasting \cite{Alexiadis2015}, to accelerate the convergence of the Monte Carlo simulations in transfer capability analysis \cite{Ramezani2009}, and to study the influence of power flows on the damping of critical oscillatory modes \cite{Rueda2009}.

\subsubsection{k-means Algorithm}
Among many data clustering methods, k-means algorithm is one of the most often used methods for clustering. This method is very simple and especially suitable for large data sets and can be easily implemented in solving many practical problems.

For a given data set \(\mathcal{X}= \{ x_i \mid x_i\in \mathbb{R}^n,i=1,2,...,n \} \), the algorithm partitions the data into \(k \) clusters, \(C_1,C_2,...,C_k \), where \(c_1,c_2,...,c_k \) are cluster centroids or cluster means, defined as:
\begin{equation}
	{c_j} = \frac{1}{{{N_j}}}\sum\limits_{x \in {C_j}} x \tag{1} \label{eq:1},
\end{equation}
where \(N_j \) is the number of data points in cluster \(j \). Conventionally, \(k \) is an input parameter to the algorithm. The similarity of the data in a cluster is defined as their Euclidean distance to the cluster centroid. In Cartesian coordinates, the Euclidean distance between two points \(x_i\) and \(x_j\) is defined as:
\begin{equation}
	d(x_i,x_j) = \sqrt {\sum\limits_{d = 1}^n {{w_d}{{({x_{id}} - {x_{jd}})}^2}} } \tag{2} \label{eq:2},
\end{equation}
where feature weights \( w_d \) are set to $1$ in the conventional k-means algorithm. \(d\) denotes the dimensionality or the feature.

The k-means algorithm can be cast as an optimization problem with the following objective:
\begin{equation}
\mathop {\arg \min }\limits_C  \sum_{i=1}^{k} \sum_{x \in C_i} \left\| x - c_i \right\| ^2 \tag{3} \label{eq:3}
\end{equation} 

This is a NP-hard problem, several efficient heuristic solution techniques have been proposed \cite{Xu2005}. It is efficient in clustering large data sets, however being a non-convex problem, it often terminates in local optima. 

\subsubsection{Particle Swarm Optimization (PSO)}
PSO is a population-based stochastic search process used to solve global optimization problems where conventional mathematical programming approaches fail \cite{Clerc2002}. In the PSO, a swarm consists of a number of potential solutions to the optimization problem, where each particle of the swarm corresponds to a potential solution. In the context of clustering, a single particle represents a group of cluster centroids. The aim of the PSO is to find the position of a particle that results in the best evaluation of a given objective function, in our case the sum of the mean squared error (SMSE) defined as:
\begin{equation}
J_e = \dfrac{1}{N_c}\sum\nolimits_{j = 1}^{{N_c}} {\left[ \dfrac{1}{\left| C_j \right|} \sum\limits_{{x_i} \in {C_j}} {d({x_i},{c_j})}  \right]},  \tag{4} \label{eq:4}
\end{equation}
where \(N_c\) is the size of the cluster centroid vector, \(c_j\) is a cluster centroid defined in \eqref{eq:1}, \(|C_j|\) is the number of data vectors belonging to cluster \(C_j\), and \(d(\cdot)\) is the Euclidean distance defined in \eqref{eq:2}.
To search for the best solution in a multi-dimensional space, the particles `fly' through the space with different speeds and directions. In the searching process, the fitness \eqref{eq:4} of each particle is evaluated and stored. The historical best position of each particle \(p_{\mathrm{best}}\) and the global best position \(g_{\mathrm{best}}\) among all the particles are used to adjust the flying speed and the direction of the particles. 

The velocity of each particle is updated according to:
\[\begin{array}{c}
{v_i}(n + 1) = w \cdot {v_i}(n) + {c_1} \cdot \mathop{\operatorname{rand}}_{1}  \cdot \left( {p_{\mathrm{best}} - {p_i}(n)} \right)\\
 + {c_2} \cdot  \mathop{\operatorname{rand}}_{2} \cdot  \left( {g_{\mathrm{best}} - {p_i}(n)} \right) \tag{5} \label{eq:5}
\end{array}\]
where \(c_1 \) and \(c_2 \) are constants, \(\mathop{\operatorname{rand}}_{1} \in [0,1] \) and \(\mathop{\operatorname{rand}}_{2} \in [0,1] \) are randomly generated numbers, and $w$ is the inertia factor defined as:
\begin{equation}
w = {w_{\max }} - {n_{\mathrm{iter}}} \cdot  \frac{{{w_{\max }} - {w_{\min }}}}{{{N_{\mathrm{iter}}}}}.   \tag{6} \label{eq:6}
\end{equation}
The particles' position are iteratively updated as follows: 
\begin{equation}
p_i(n + 1) = {p_i}(n) + {v_i}(n + 1). \tag{7} \label{eq:7}
\end{equation}

In \cite{VanderMerwe,Ahmadyfard2008}, the authors have demonstrated that the combination of PSO and k-means clustering can improve the clustering performance or to some extent, overcome the weaknesses of the k-means algorithm. We build on that by proposing a novel self-adaptive PSO-k-means clustering algorithm, discussed in more detail in Section IV.B.

\vspace{-0.15cm}
\subsection{Feature Selection}
An operating condition of a power system is defined by a set of system variables, or features, e.g. generator active and reactive powers, bus voltage magnitudes and angles, load levels, etc. Feature selection is a process of selecting a subset of relevant features that is necessary and sufficient to describe the target concept by reducing the dimensionality of the input data and enhancing generalization by reducing over-fitting \cite{ROBNIK-SIKONJA2003}.
Feature selection has attracted significant attention in DSA, e.g. in \cite{Jensen2001, Mohammadi2009, Niazi2003}.

\subsubsection{Relief Algorithm}
 A popular feature selection algorithm with little application in power systems is ReliefF \cite{KIRA1992, ROBNIK-SIKONJA2003}. The main idea of the original Relief algorithm \cite{KIRA1992} is to estimate features' ability, represented by features' weights, to distinguish between instances, power system operating conditions in our case, that are near to each other.

\begin{algorithm}
\caption{RReliefF feature selection algorithm \cite{ROBNIK-SIKONJA2003}} \label{RReliefF}
\textbf{Input}: For each training instance $r \in \mathcal{R}$ a vector of attribute values $a \in \mathcal{A}$ and predicted values $\lambda \in \mathcal{L}$.\\
\textbf{Output}: For each training instance $r \in \mathcal{R}$ a vector $w \in \mathbb{R}^{\left| \mathcal{A} \right|}$ of estimations of the qualities of attributes $a \in \mathcal{A}$.
\begin{algorithmic}[1]
\State{Set all $w$ to 0};
\For{$i \gets 1, m$}
	\State{Randomly select instance $r_i$};
	\State{Select $k$ instances $q_j$ nearest to $r_i$};
		\For{$j \gets 1, k$}
			\State{$n^{\textrm{dc}} \gets n^{\textrm{dc}} + \textrm{diff} \left( \lambda(\cdot),r_i,q_j \right) \cdot  d(r_i,q_j)$}
				\For{$l \gets 1, \left| \mathcal{A} \right|$}
					\State{$n^{\textrm{da}}_l \gets n^{\textrm{da}}_l + \textrm{diff} \left( l,r_i,q_j \right) \cdot d(r_i,q_j)$}
					\State{$n^{\textrm{dca}}_l \gets n^{\textrm{dca}} + $}
					\State{$\textrm{diff} \left( \tau(\cdot),r_i,q_j \right) \cdot \textrm{diff} \left( l,r_i,q_j \right) \cdot  d(r_i,q_j)$}
				\EndFor
		\EndFor	
\EndFor
\For{$l \gets 1, a$}
	\State{$w_l \gets n^{\textrm{dca}}/n^{\textrm{dc}} - (n^{\textrm{da}} - n^{\textrm{dca}})/( m - n^{\textrm{dc}}) $}
\EndFor
\end{algorithmic}
\end{algorithm}

The original Relief algorithm \cite{KIRA1992} is limited to two class problems. Its extensions, ReliefF and RReliefF can also deal with multi-class and regression problems, respectively \cite{ROBNIK-SIKONJA2003}. The psudo code for the RReliefF algorithm used in this study is shown in Algorithm \ref{RReliefF}, where $n_{\textrm{dc}}$, $n_{\textrm{da}}$, and $n_{\textrm{dca}}$ denote the weights for the prediction values of different prediction (line 6), different attribute (lines 8) and for different prediction and different attribute (line 9 and 10), respectively.
The term $d(r_i,q_j)$ takes into account the distance between the two instances $r_i$ and $q_j$. It is defined as:
\begin{equation}
	d(r_i,q_j) = \dfrac{d_1(r_i,q_j)}{\sum_{l=1}^k d_1(r_i,q_j)} \tag{8} \label{eq:8}
\end{equation} 

\noindent
Closer instances should have greater influence, so the influence of instance $r_j$ is exponentially decreased with the distance from the given instance $r_i$:
\begin{equation}
	d_1(r_i,q_j) = e^{-\left( \textrm{rank} (r_i,q_j)/\sigma \right)^2} \tag{9} \label{eq:9}
\end{equation} 

\noindent
where $\textrm{rank} (r_i,q_j)$ is the rank of the instance $q_j$ in a sequence of instances ordered by the distance from $r_i$ and $\sigma$ is a user defined parameter controlling the influence of the distance.

\vspace{-0.15cm}
\section{A Novel Fast Stability Scanning Framework}
In the original simulation platform \cite{Marzooghi2014}, stability analysis is performed on all operating points, which is time consuming. We propose a framework for fast stability scanning to achieve a significant computational speed-up. The framework consists of three parts: (i) feature selection, (ii) clustering, and (iii) stability analysis. The pseudo code of the framework is shown in Algorithm \ref{fast_scanning}.

\begin{mydef}
Let $\mathcal{R} = \{r_i\mid r_i \in \mathbb{R}^{ \left| \mathcal{A} \right|}, i = 1,2,\ldots, \left| \mathcal{R} \right| \}$ denote a steady-state power system operating condition, uniquely defined by a set of attributes $\mathcal{A} = \{a_i\mid r(a_i) \in [-1,1]^{\left| \mathcal{R} \right|}, i = 1,2,\ldots, \left| \mathcal{A} \right| \}$, where $r(a_i)$ is a normalized numerical value of attribute $a_i$ across all operating conditions. For each operating condition $r_i \in \mathcal{R}$, we compute a stability index $\lambda_i \in \mathbb{R}$. The task of \textbf{fast stability scanning}  is to cluster $\mathcal{R}$  into a set of representative clusters $\mathcal{C}$ represented by cluster centroids $c \in \mathbb{R}^{|\mathcal{A}|} $, so that $\left| \mathcal{C} \right| <  \left| \mathcal{R} \right|$, and to compute a stability index $\hat{\lambda}$ using cluster centroids $c \in \mathcal{C}$, so that $| \lambda - \hat{\lambda} | \le \epsilon$ for all $r \in \mathcal{R}$, where $\epsilon$ is a predefined tolerance.
\end{mydef}

\begin{algorithm}
\caption{Fast stability scanning framework.} \label{fast_scanning}
\textbf{Input}: Set of operating conditions $\mathcal{R}$, feature selection performance $\rho$ and tolerance $\epsilon_{\mathrm{f}}$, set of features $\mathcal{A}$.\\
\textbf{Output}: Stability index $\lambda$ for each for each $r \in \mathcal{R}$, minimum cluster distance $\epsilon_{\mathrm{c}}$, minimum data distance $\epsilon_{\mathrm{d}}$.
\begin{algorithmic}[1]
\While{$\rho \geq \epsilon$}
	\State{Randomly select a training instance $r_i$;}
	\State{Run feature selection using RReliefF (Algorithm \ref{RReliefF});}
	\State{Update feature weights for all $a \in \mathcal{A}$ (\ref{eq:10});}
\EndWhile
\State{Run self-adaptive PSO-k-means clustering (Algorithm \ref{PSO-k-means});}
\For{$c \gets 1, \left| \mathcal{C} \right|$}
	\State{Calculate $\lambda(c)$;}
\EndFor
\For{$r \gets 1, \left| \mathcal{R} \right|$}
	\State{Assign $\lambda(c)$ to $r(c)$;}
\EndFor
\end{algorithmic}
\end{algorithm} 

A time-series analysis of one full year with an hourly resolution results in $\left| \mathcal{R} \right| = 8760$. A minimum feasible set $\mathcal{A}$ includes voltage magnitudes and angles at all buses in the system, active and reactive demands, and active and reactive powers of all generators in the system. Without the loss of generality, however, $\mathcal{A}$ can also include derived variables, such as transmission line flows.

The framework proposed in this paper bears similarities and differences with online DSA. They both involve knowledge base generation and
feature selection. The first difference is in the offline simulation: DSA requires a big knowledge base to achieve high accuracy mapping as a supervised learning method, while fast scanning involves much smaller simulation for feature selection and represented operating points stability analysis as an unsupervised method. The second difference is in the application: DSA is an operational tool which requires fast mapping of current or imminent operating conditions and very high accuracy since the mapping result is the basis for preventive or emergency control, while fast scanning is developed as a planning tool which aims to scan large amount of scenarios across long horizons and to provide planners with the stability level of the system under study.

\vspace{-0.15cm}
\subsection{Novel Feature Selection (Lines 1-5 in Algorithm \ref{fast_scanning})}
Compared to conventional DSA, we propose two innovations in feature selection: (i) both feature ranks and weights are used in clustering, and (ii) the size of the required training set for feature selection is determined adaptively to reduce the simulation time. 

In this paper, the candidate features considered for clustering include active and reactive powers loads, and reactive powers loads of thirteen synchronous generators including one CSP, six wind farms and two utility PV farms, HVDC links' active and reactive powers, and inter-area active and reactive power flows.
In \cite{Zhou2010, Shayanfar2012, Xu2012, Amjady2007}, feature ranks are used to select a subset of candidate features used in the classifier that determines the feature weights. In this study, both feature ranks and weights are used for clustering. This requires preprocessing, due to two reasons: (i) the accumulated effect of many unimportant features may mask the effect of a smaller number of dominant features, and (ii) to improve the representativeness of the cluster centroids', the degree of segmentation for features with large variance should be increased.
We propose the following weight adjustment for all $a \in \mathcal{A}$:

\begin{equation}
	\tilde{w_i} = C \cdot w_i \cdot \dfrac{{\textrm{var}(r(a_i))}}{{\log \left(2 \cdot \textrm{rank}(a_i) \right) }} \tag{10} \label{eq:10}
\end{equation} 

\noindent
where $C$ is a tunable parameter, and $\tilde{w_i}$ and $w_i$ are adjusted and original feature weights, respectively.

\vspace{-0.15cm}
\subsection{Self-adaptive PSO-k-means Clustering (Line 6 in Algorithm \ref{fast_scanning})}
The conventional k-means clustering algorithm has two inherent drawbacks: (1) its clustering performance depends on randomly assigned initial cluster centroids, which can lead to unreliability; (2) the algorithm is based on gradient descent and can thus easily terminate in local optima. 
In the PSO-k-means algorithm, the solution of the PSO can be used as the initial k-means cluster centroids, which can avoid the algorithm trapping in local optima.  However, like any other global optimization algorithm, the PSO is prone to premature convergence. This may be improved by increasing the size of the swarm but at the cost of an increased computational burden. Another issue is to determine the cluster numbers and how to deal with empty clusters.  To address these issues, we propose a self-adaptive PSO-k-means clustering algorithm, described in Algorithm \ref{PSO-k-means}.

\begin{algorithm}
\caption{Self-adaptive PSO-k-means clustering.} \label{PSO-k-means}
\textbf{Input}: PSO iteration limit $\mathrm{MaxIter}$\\
\textbf{Output}: Cluster centroids $C$.
\begin{algorithmic}[1] 
\State{Initialize $C_{0}$, $V_{0}$, $p_{\mathrm{best},0}$, $g_{\mathrm{best},0}$;}  
\While{iteration $\leq \mathrm{MaxIter}$}
	\For{$i \gets 1, \mathrm{SwarmSize}$}
	   \State{Update $V_i, C_i$ (\ref{eq:5});}
		 \State{Update $p_{\mathrm{best}, i},g_{\mathrm{best},i}$ if required;}
		 \State{Search space limit check;}
	\EndFor
	\State{Calculate swarm fitness variance (\ref{eq:4});}	
	\State{Calculate mutation probability $p_\mathrm{m}$ \cite{LvZhensu2004};}	
		 \If{$p_\mathrm{m} > \mathop{\operatorname{rand}} \in [0,1]$}
		    \State{Mutate $g_{\mathrm{best}}$ (\ref{eq:9});} 
		 \EndIf
\EndWhile
\State{The best particle position is used as initial cluster centroids for k-means;}
  \Repeat
	  \State{Perform k-means clustering;}
  	\State{Remove empty clusters;}
  	\State{Create new cluster for data points $d(r,c(r)) > \epsilon_{\mathrm{d}}$;}
  	\State{Combine clusters if $d(c_i,c_j) < \epsilon_{\mathrm{c}}$;}
  \Until {convergence}
\end{algorithmic}
\end{algorithm} 

The algorithm starts with the initialization of the PSO particles. Random cluster centroids (operating points in our case) are assigned as the particles' initial position \( C_{0}\), and local best \(p_{\mathrm{best},0}\), global best \(g_{\mathrm{best},0}\) are calculated using a random initial velocity \(V_{0}\). 
The PSO (Lines 2 to 13) is ran first to locate the best initial position, which is then used by the k-means clustering in the second stage (Line 14 to 20).
In the PSO run, the position and the direction of each particle are updated in every iteration. 
The issue with the conventional PSO algorithm is that a particle may fly out of the load-flow solution space, resulting in a divergent load flow and hence an infeasible cluster. To overcome this, the nearest feasible position within the solution space is used instead of the invalid position (Line 6). To the premature convergence of the conventional PSO algorithm, we adopt a technique proposed in \cite{LvZhensu2004} that monitors the fitness variance of all the particles in the swarm in each iteration and uses it as an indicator of premature convergence. A mutation probability \(p_\mathrm{m}\) is calculated according to \cite{LvZhensu2004} and used as a trigger for a mutation of \(g_{\mathrm{best}}\) (Line 10 to 12). The mutation of \(g_{\mathrm{best}}\) is defined as:
\begin{equation}
g_{\mathrm{best},k} = g_{\mathrm{best},k} \cdot \left(1 + \frac{\eta}{2} \right), \tag{10} \label{eq:11}
\end{equation}
where \(\eta\) is a normally distributed random variable.

\subsection{Stability Scanning (Lines 7-11 in Algorithm \ref{fast_scanning})}
Compared with the initial number of operating points, the number of representative clusters resulting from clustering is much smaller. The stability analysis is performed on cluster centroids using conventional stability analysis. The stability index $\lambda(c)$ is assigned to every operating point $r(c)$ represented by the cluster centroid $c$.
Given $\left| \mathcal{C} \right| <  \left| \mathcal{R} \right|$, the computational time is significantly reduced.

\section{Simulation Results}
In order to evaluate the efficacy of the proposed fast stability scanning framework, we performed small signal stability (SSA) and steady-state voltage stability analysis (VSA) of a simplified model of the NEM in the year 2030 described in Section III. Fast stability scanning is performed using the representative cluster centroids and the results are compared with the time-consuming time-series stability analysis, that uses all 8760 operating points.
For small-signal stability, the damping ratio of the inter-area oscillation mode between Areas 2 and Area 4 is used as the stability index, whereas for voltage stability, we used the loading margin assuming a uniform load increase at all load buses in the system, where all generators increase their production in proportion to the base case. We first present the results of feature selection and clustering, followed by the results of stability scanning.

\subsection{Feature Selection}
Tables \ref{TableSSA} and \ref{TableVSA} show the initial weights and ranks and adjusted weights and ranks for SSA and VSA, respectively. 

\begin{table}
\centering
\caption{Features and weights for SSA}
\label{TableSSA}
\begin{tabular}{|c|c|c|c|c|}
\hline
\textbf{Feature name} & \textbf{\begin{tabular}[c]{@{}c@{}}Initial\\  weights\end{tabular}} & \textbf{\begin{tabular}[c]{@{}c@{}}Initial\\  rank\end{tabular}} & \textbf{\begin{tabular}[c]{@{}c@{}}Adjusted\\  weights\end{tabular}} & \textbf{\begin{tabular}[c]{@{}c@{}}Adjusted\\  rank\end{tabular}} \\ \hline
Sync11\_P     & 0.106      & 1           & 7.709         & 1  \\ \hline
WF04\_P       & 0.098      & 2           & 2.267         & 2  \\ \hline
Sync11\_Q     & 0.062      & 3           & 1.748         & 3  \\ \hline
PV02\_P       & 0.054      & 4           & 1.020         & 4  \\ \hline
PV01\_Q       & 0.041      & 8           & 0.538         & 5  \\ \hline
PV01\_P       & 0.040      & 11          & 0.465         & 6  \\ \hline
WF04\_Q       & 0.039      & 12          & 0.397         & 7  \\ \hline
PV02\_Q       & 0.035      & 17          & 0.396         & 8  \\ \hline
Sync09\_P     & 0.048      & 5           & 0.310         & 9  \\ \hline
Sync08\_Q     & 0.047      & 6           & 0.297         & 10 \\ \hline
\end{tabular}
\end{table}

\begin{table}
\centering
\caption{Features and weights for VSA}
\label{TableVSA}
\begin{tabular}{|c|c|c|c|c|}
\hline
\textbf{Feature name} & \textbf{\begin{tabular}[c]{@{}c@{}}Initial \\ weights\end{tabular}} & \textbf{\begin{tabular}[c]{@{}c@{}}Initial \\ rank\end{tabular}} & \textbf{\begin{tabular}[c]{@{}c@{}}Adjusted \\ weights\end{tabular}} & \textbf{\begin{tabular}[c]{@{}c@{}}Adjusted \\ rank\end{tabular}} \\ \hline
WF06\_P   & 0.130 & 1   & 7.416 & 1  \\ \hline
WF05\_P   & 0.116 & 3   & 2.372 & 2  \\ \hline
WF06\_Q   & 0.109 & 4   & 2.066 & 3  \\ \hline
Inter-P3  & 0.128 & 2   & 1.973 & 4  \\ \hline
HVDC3S\_Q & 0.096 & 5   & 1.110 & 5  \\ \hline
WF02\_P   & 0.068 & 6   & 0.704 & 6  \\ \hline
WF05\_Q   & 0.036 & 7   & 0.506 & 7  \\ \hline
WF03\_P   & 0.023 & 11  & 0.299 & 8  \\ \hline
WF02\_Q   & 0.027 & 8   & 0.250 & 9  \\ \hline
Inter-P2  & 0.025 & 10  & 0.145 & 10 \\ \hline
\end{tabular}
\end{table}

The results confirmed the necessity of the feature selection before clustering, showing the features weights resulting from the feature selection are quite different, which reflects the different features' impact on SSA and VSA. It is interesting to observe that the generator Sync11 (CSP) and Wind Farm 04, both located in northern QLD, have a significant impact on the oscillation mode  between Areas 2 and 4.

In order to find the dominant features, the size of the training set is progressively increased by randomly picking the operating points from the time-series analysis until the resulted feature ranks and weights converge. Compared to conventional DSA where the size of the training set for feature selection is fixed, our approach avoids unnecessary computation thus reducing the computational burden, and also prevents overfitting.

\begin{figure}
\centerline{\includegraphics[height=9cm]{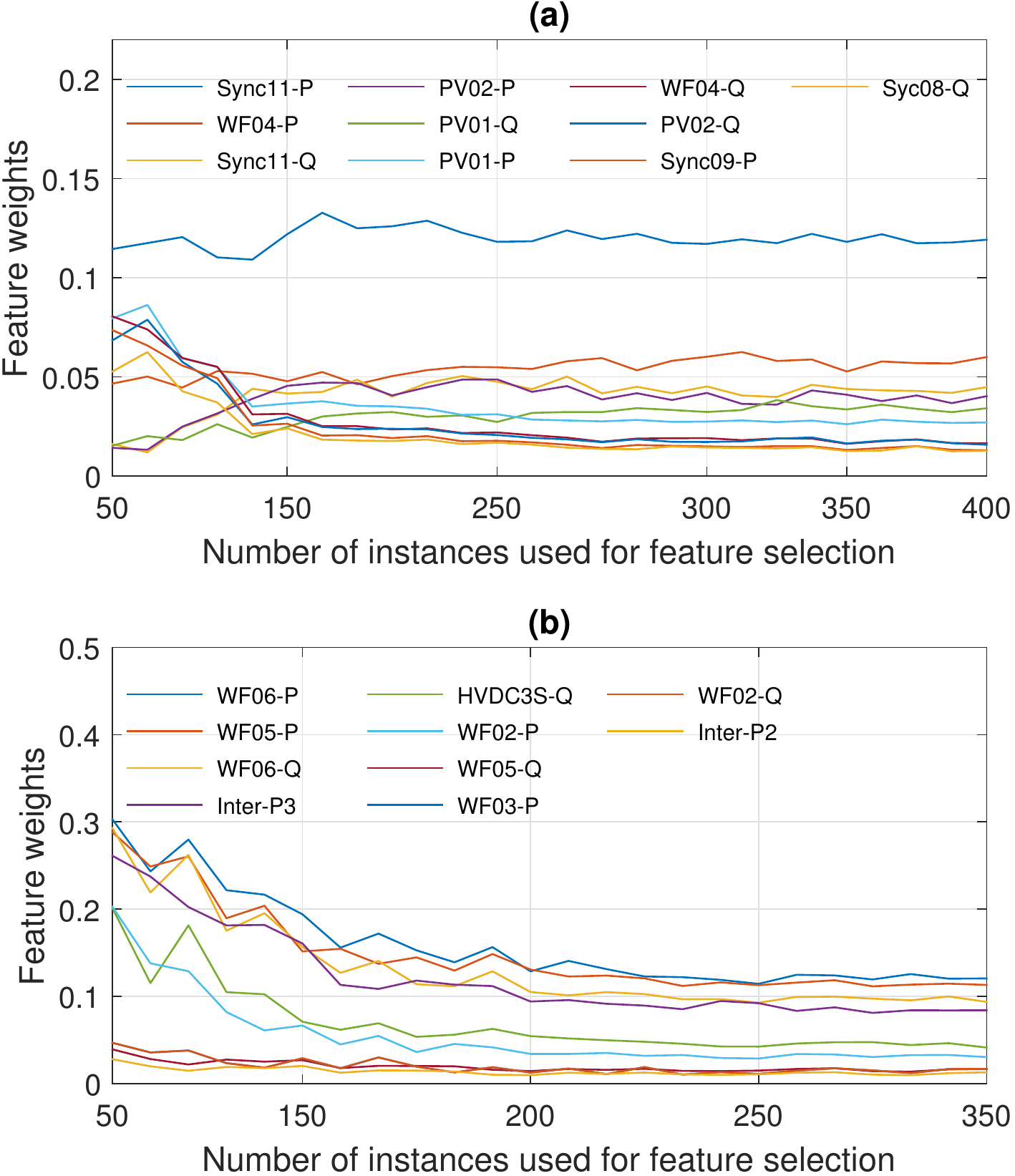}}
\caption{Convergence of feature selection: (a) SSA, (b) VSA.}
\label{fig:2}
\end{figure}

Fig. \ref{fig:2} shows the convergence process. Observe that a sufficient accuracy is achieved after 300 iterations. Note that the stability index need to be calculated using conventional methods for all operating points used for feature selection.

\subsection{Clustering}
Self-adaptive PSO-k-means weighted clustering is used to find typical generation-load patterns. Clustering reduces the number of data points from 8760 operating points resulting from the time-series analysis to 555 and 421 clusters, for SSA and VSA, respectively, which represents a dimensionality reduction of 95.2\% and 93.7\%, respectively.

\begin{figure}
\centerline{\includegraphics[height=4.5cm]{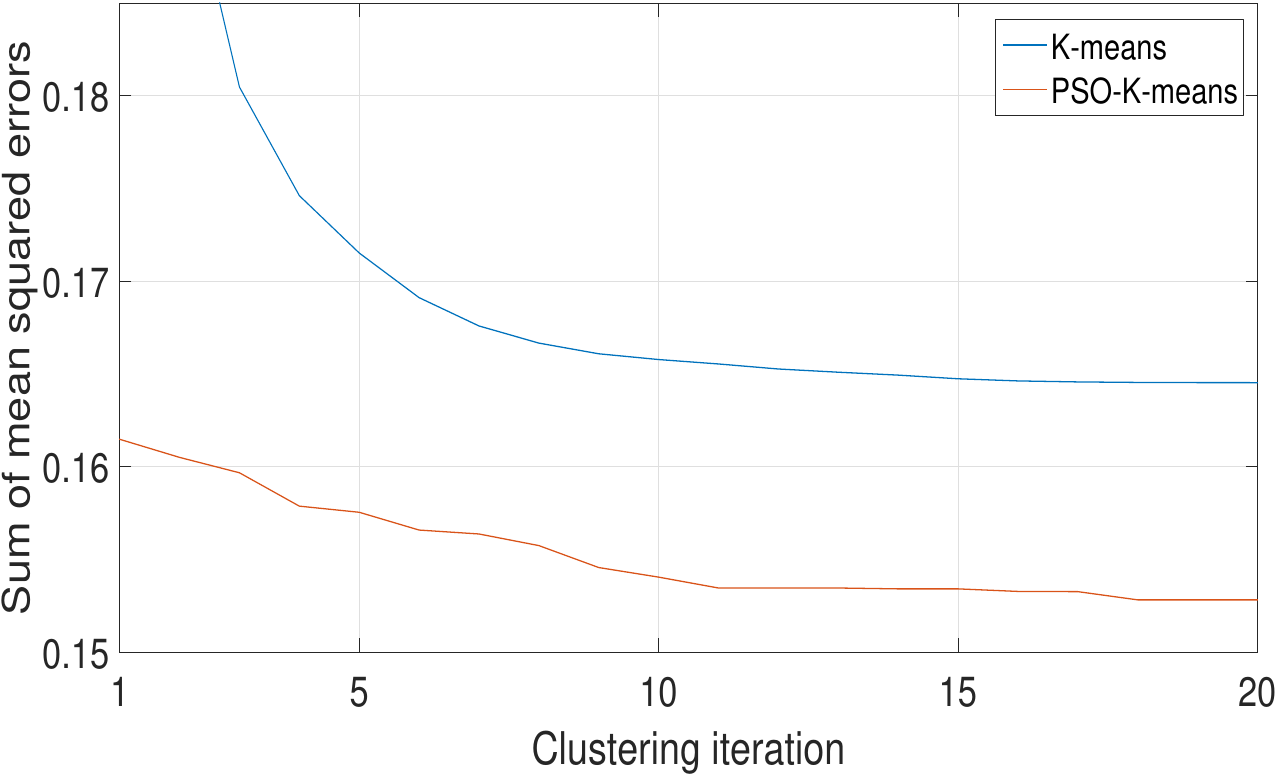}}
\caption{Comparison of the clustering results: conventional k-means vs. the proposed self-adaptive PSO-k-means.}
\label{fig:3}
\end{figure}

Fig. \ref{fig:3} compares the clustering results using the conventional k-means and the proposed self-adaptive PSO-k-means algorithm. Observe that the k-means algorithm starts from a randomly assigned cluster centroid that is normally far away from the global optimum. Therefore, the SMSE of the k-means is much larger than the PSO-k-means SMSE in the first a few iterations. The PSO-k-means, on the other hand, starts with a much smaller SMSE, and has a better performance overall.

\subsection{Small-signal stability}
\begin{figure}
\centerline{\includegraphics[height=9cm]{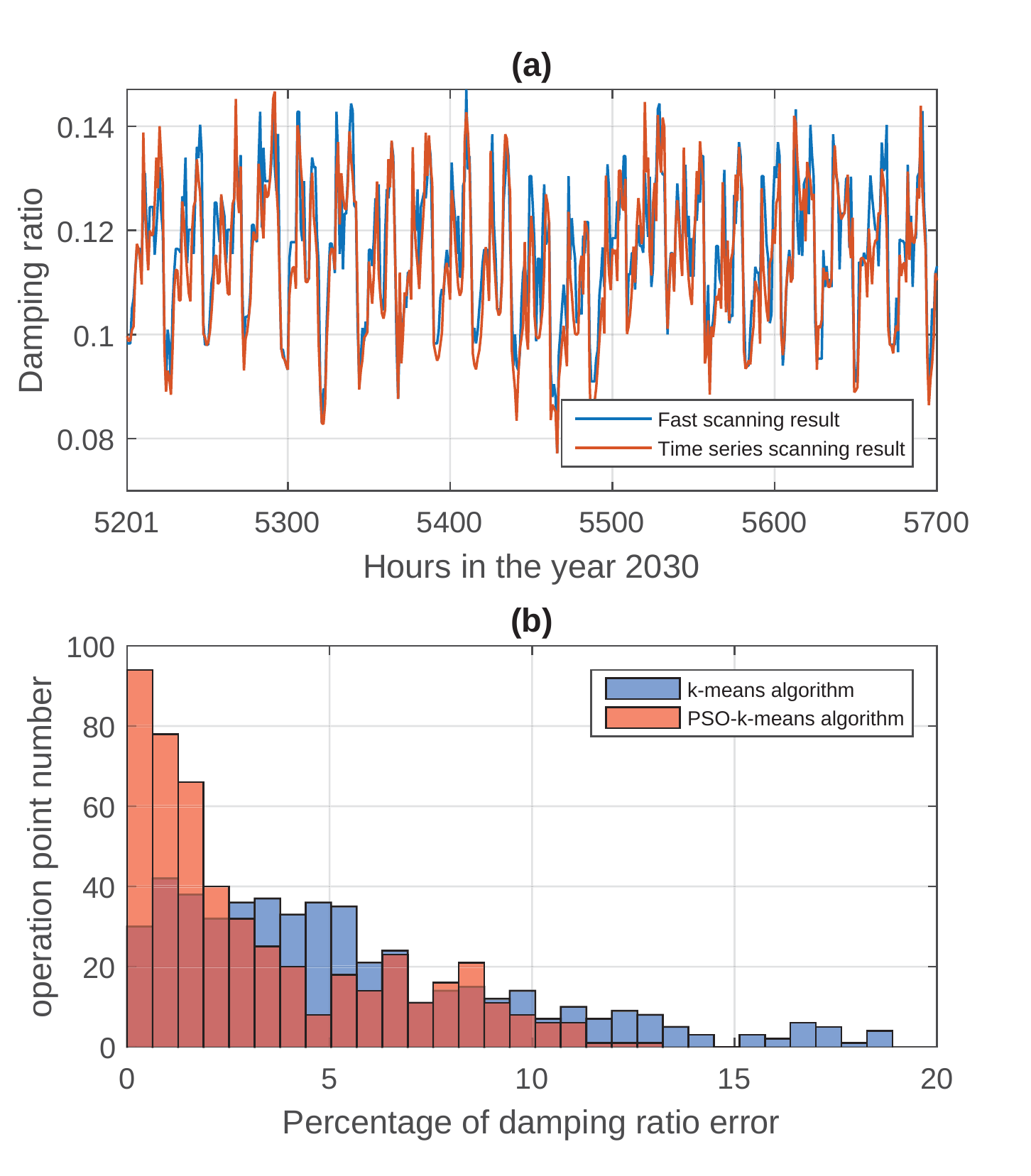}}
\caption{SSA critical damping ratio fast scanning results: (a)  time series, (b) error distribution using PSO-k-means and k-means.}
\label{fig:4}
\end{figure}

For the sake of illustration, a section of the damping ratio of the inter-area oscillation mode between Areas 2 and 5 between hours 5201 and 5700 is shown in Fig. \ref{fig:4} (a), which reveals a close agreement between the fast scanning results and the time series analysis.
To verify that statistically, the damping ratios were calculated for 500 randomly selected operating conditions and compared with the values obtained from fast stability scanning. 
Fig. \ref{fig:4} (b) compares the error distribution of the damping ratio as result of fast scanning using the conventional k-means (blue bins) and the proposed PSO-k-Means algorithm (red bins). 
Observe that the error the proposed PSO-k-means algorithm is kept below 14\%, with the highest density in the 0-4\% range, while for the conventional k-means, the error can be as high as 19\%. The average percentage error is 3\% to 5\% for PSO-k-means and k-means, respectively.

\subsection{Voltage stability}
\begin{figure}
\centerline{\includegraphics[height=9cm]{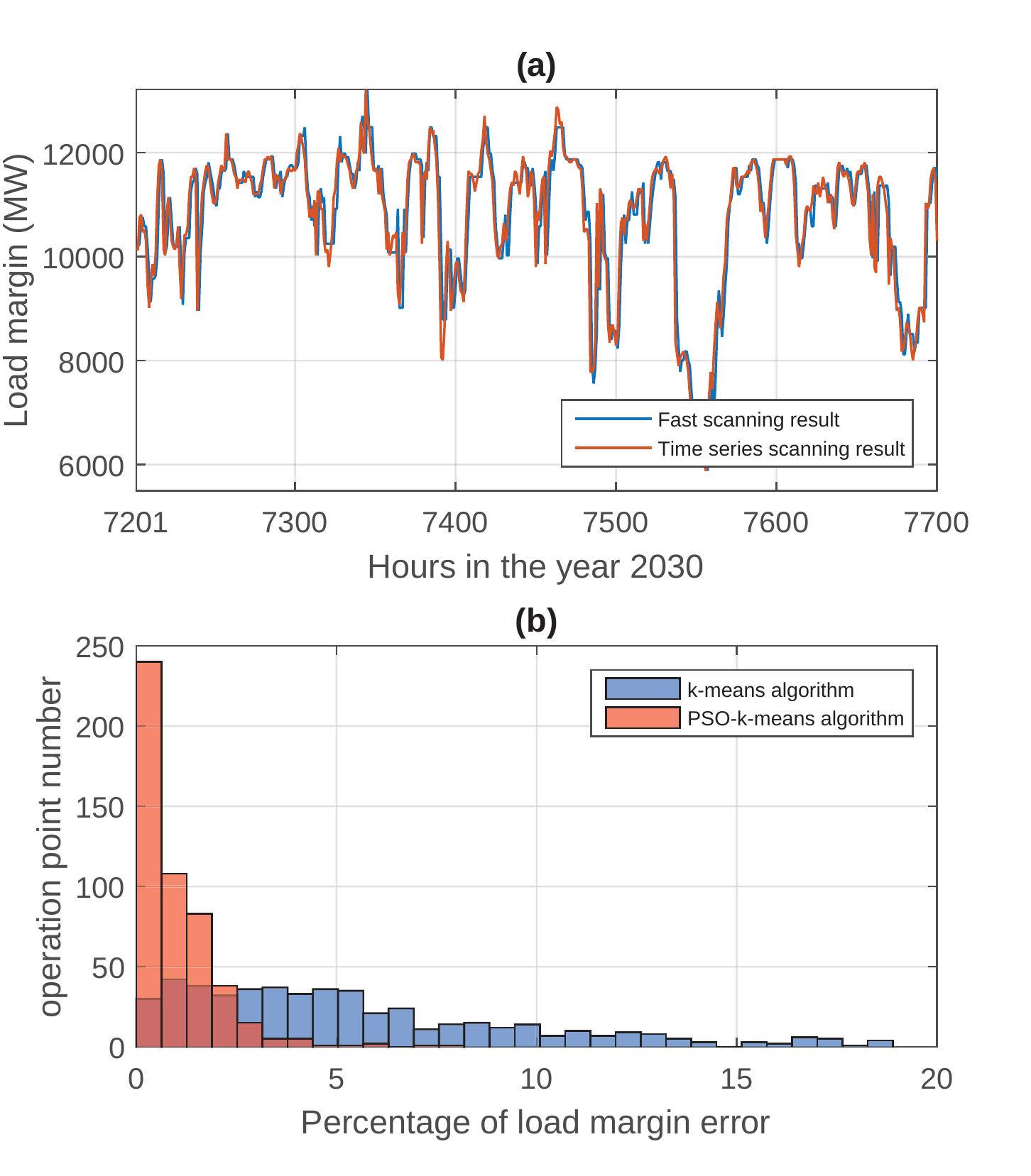}}
\caption{VSA loading margin fast scanning results: (a) time series, (b) error distribution using PSO-k-means and k-means.}
\label{fig:5}
\end{figure}

To illustrate the performance of fast stability scanning for voltage stability analysis, Fig. \ref{fig:5} (a) shows the loading margin between hours 7201 and 7700. Again, in order to verify the fast scanning accuracy, we calculated the loading margin for 500 randomly selected operating conditions and compared the results with the values obtained with fast stability scanning. Fig. \ref{fig:5} (b) compares the error distribution of the load margin using the proposed PSO-k-means (red bins) and the conventional k-means (blue bins). Observe that the error is mostly kept below 4\%, with the highest density in the 0-2\% range for the PSO-k-means.
Similar to the small-signal stability, the proposed PSO-k-means algorithm performs much better. In this case, the average percentage error decreases from 6\% to 1\% and the maximum error decreases from 19\% to 8\% compared to the conventional k-means.

\subsection{Worst case operating point shift}

\begin{figure}
\centerline{\includegraphics[height=4.5cm]{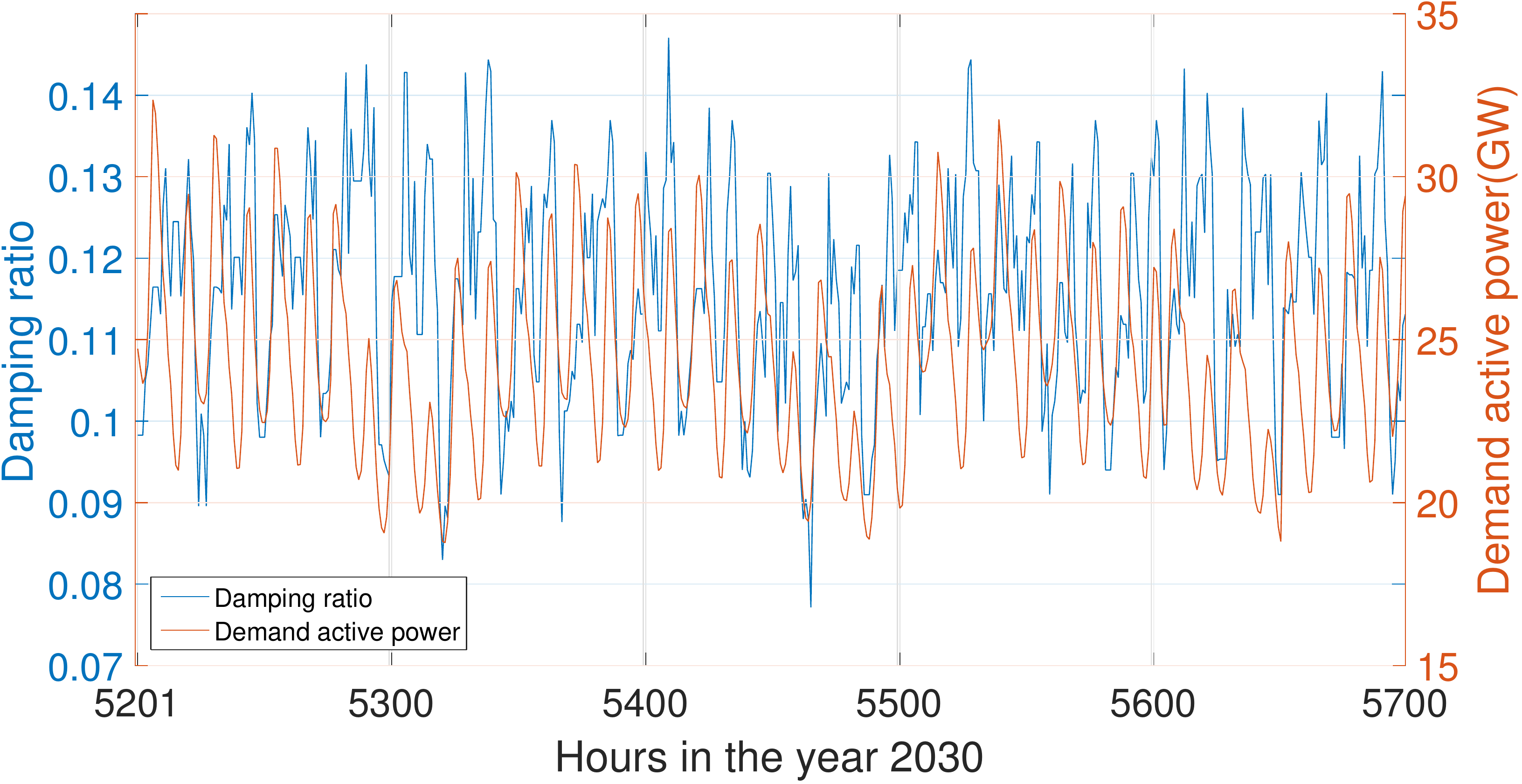}}
\caption{SSA: Critical mode damping ratio vs. demand.}
\label{fig:6}
\end{figure}

Conventionally in power system planning, worst case conditions are considered when the system is the most stressed, and stability studies are conducted under these conditions. In order to clearly see the relationship between the critical damping ratio and the system generation/demand level, with the constructed inter-area oscillation mode damping ratio trace, the minimum damping ratio happens at hour 5466 in the year 2030.  In Fig. \ref{fig:6}, the damping ratio trace between hour 5201 and 5700 is given, total demand in NEM of the same time slot is compared with the damping ratio. It can be observed that the minimum damping ratio does not coincide with the local maximum load level, nor the maximum load level in the year 2030, the observation of the worst case point shifting is in accordance with \cite{Vittal2010}.

\begin{figure}
\centerline{\includegraphics[height=4.5cm]{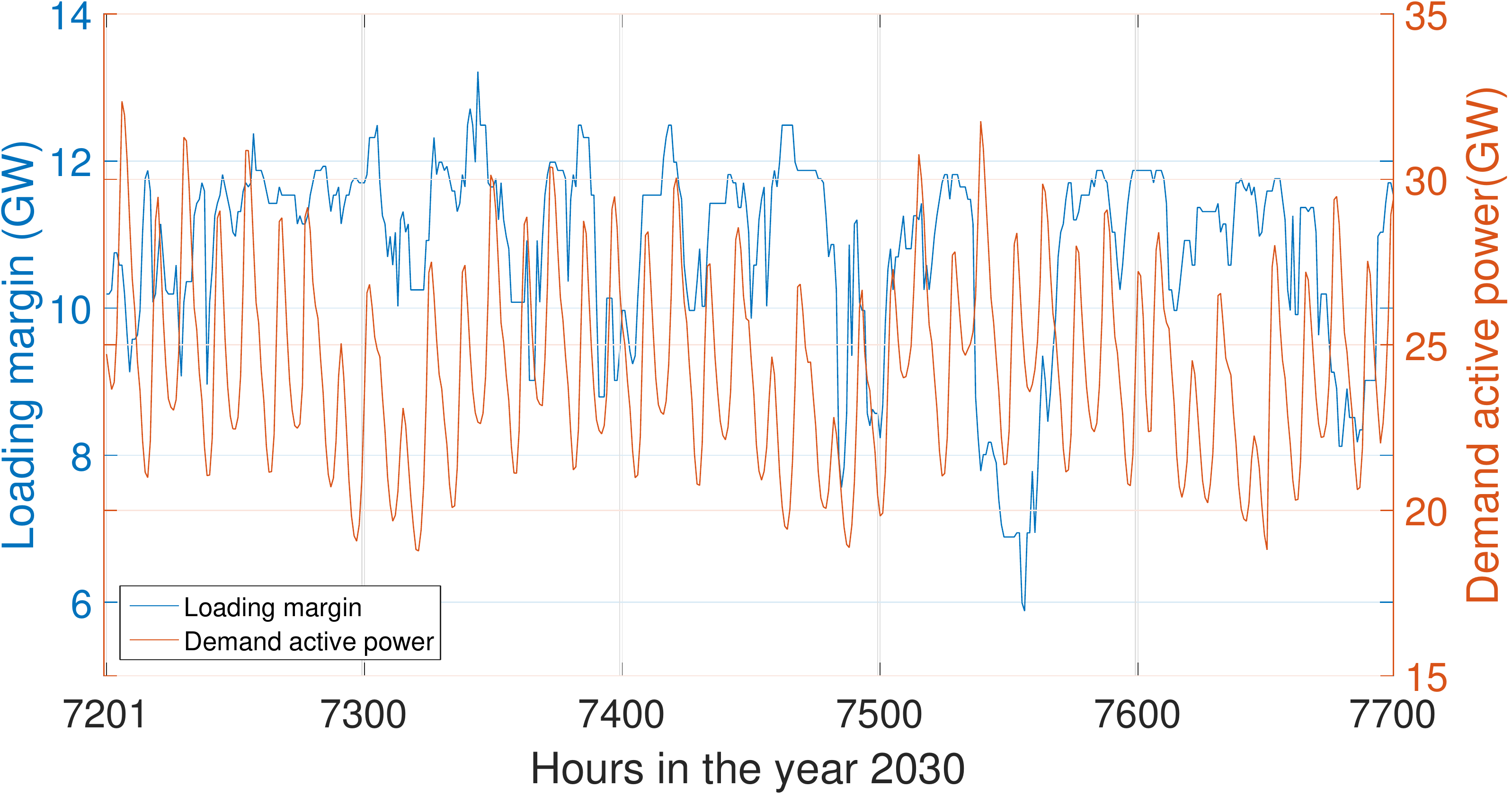}}
\caption{VSA: Loading margin vs. demand.}
\label{fig:7}
\end{figure}

Similarly, we plotted the loading margin and the total system demand for a period of 500 hours in Fig. \ref{fig:7}. Observe that there is little correlation between high/low demand level and the low/high loading margin, which justifies the time series approach compared to a conventional approach where only a small number of the most critical conditions is analyzed.

\subsection{Simulation burden of stability scanning}
The simulations were performed on a 64-bit Xeon 2.60GHz workstation with 256GB RAM. Compared to full-time series stability analysis, the computational burden has been reduced from 220 min to 21 min and from 960 min to 90 min, for SSA and VSA, respectively, which represents about a ten-fold reduction with a satisfactory accuracy. It is observed that the feature selection (30 seconds) and the clustering (5 minutes) computation does not affect the reduction of computation much. 

\section{Conclusion}
Unlike the conventional power system planning that aims to find the optimal transmission and/or generation expansion plan, the future grid analysis considers scenarios that are not mere extrapolations of the existing grid. Next, to capture the intra-seasonal variation in the RES output, we need to use time series analysis as opposed to picking a small number of the most critical operating condition, as it is done conventionally. The challenge of future grid stability analysis is the sheer number of operating conditions that need to be analyzed. In this paper, we have proposed a novel framework for fast stability scanning of future grids scenarios. The framework is based on a novel feature selection algorithm that makes it possible to perform clustering using both feature ranks and weights. To reduce the number of clusters, we proposed a novel self-adaptive PSO-k-means clustering technique that determines the optimal cluster number. The case study demonstrated the suitability of the proposed framework. Considering the level of detail required for future grid analysis, an acceptable accuracy is achieved with a more than a ten-fold speed-up.

\bibliographystyle{IEEEtran}
\bibliography{Fast_Scanning_Journal_arXiv}

\begin{IEEEbiography}[{\includegraphics[width=1in,height=1.25in,clip,keepaspectratio]{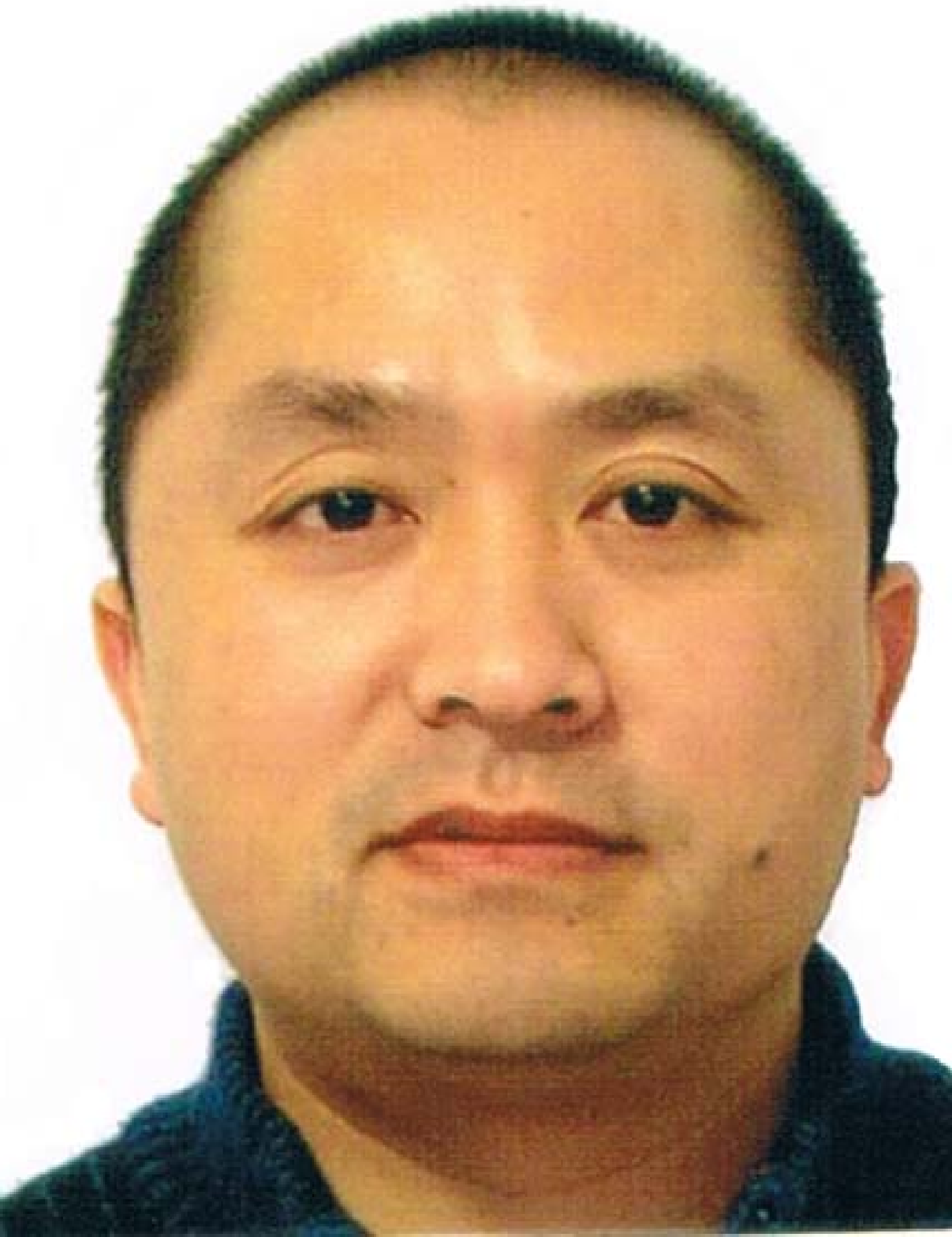}}]{Ruidong Liu} received the B.E. in power engineering from the University of Sydney, NSW, Australia and the M.E. in electrical engineering from the HoHai University, Nanjing, China. He is now pursuing the Ph.D. degree at the University of Sydney, NSW, Australia.
His research interests include power system stability and control, power system planning, power system protection and machine learning applications in power engineering.
\end{IEEEbiography}

\begin{IEEEbiography}[{\includegraphics[width=1in,height=1.25in,clip,keepaspectratio]{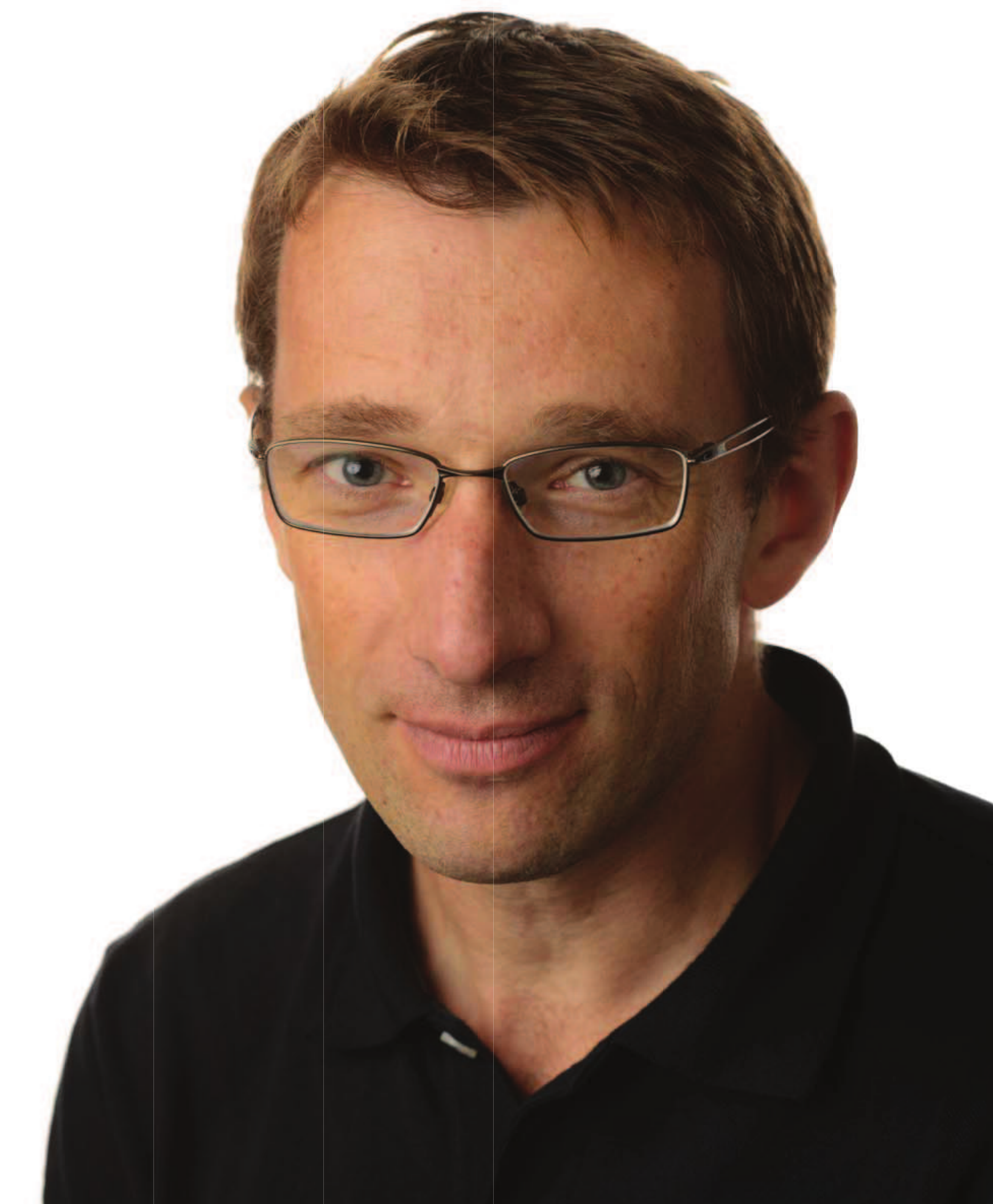}}]{Gregor Verbi\v{c}} (S’98, M’03, SM’10) received the B.Sc., M.Sc., and Ph.D. degrees in electrical engineering from the University of Ljubljana, Ljubljana, Slovenia, in 1995, 2000, and 2003, respectively. In 2005, he was a NATO-NSERC Postdoctoral Fellow with the University of Waterloo, Waterloo, ON, Canada. Since 2010, he has been with the School of Electrical and Information Engineering, The University of Sydney, Sydney, NSW, Australia. His expertise is in power system operation, stability and control, and electricity markets. His current research interests include integration of renewable energies into power systems and markets, optimization and control of distributed energy resources, demand response, and energy management in residential buildings. He was a recipient of the IEEE Power and Energy Society Prize Paper Award in 2006. He is an Associate Editor of the IEEE Transactions on Smart Grid.
\end{IEEEbiography}

\begin{IEEEbiography}[{\includegraphics[width=1in,height=1.25in,clip,keepaspectratio]{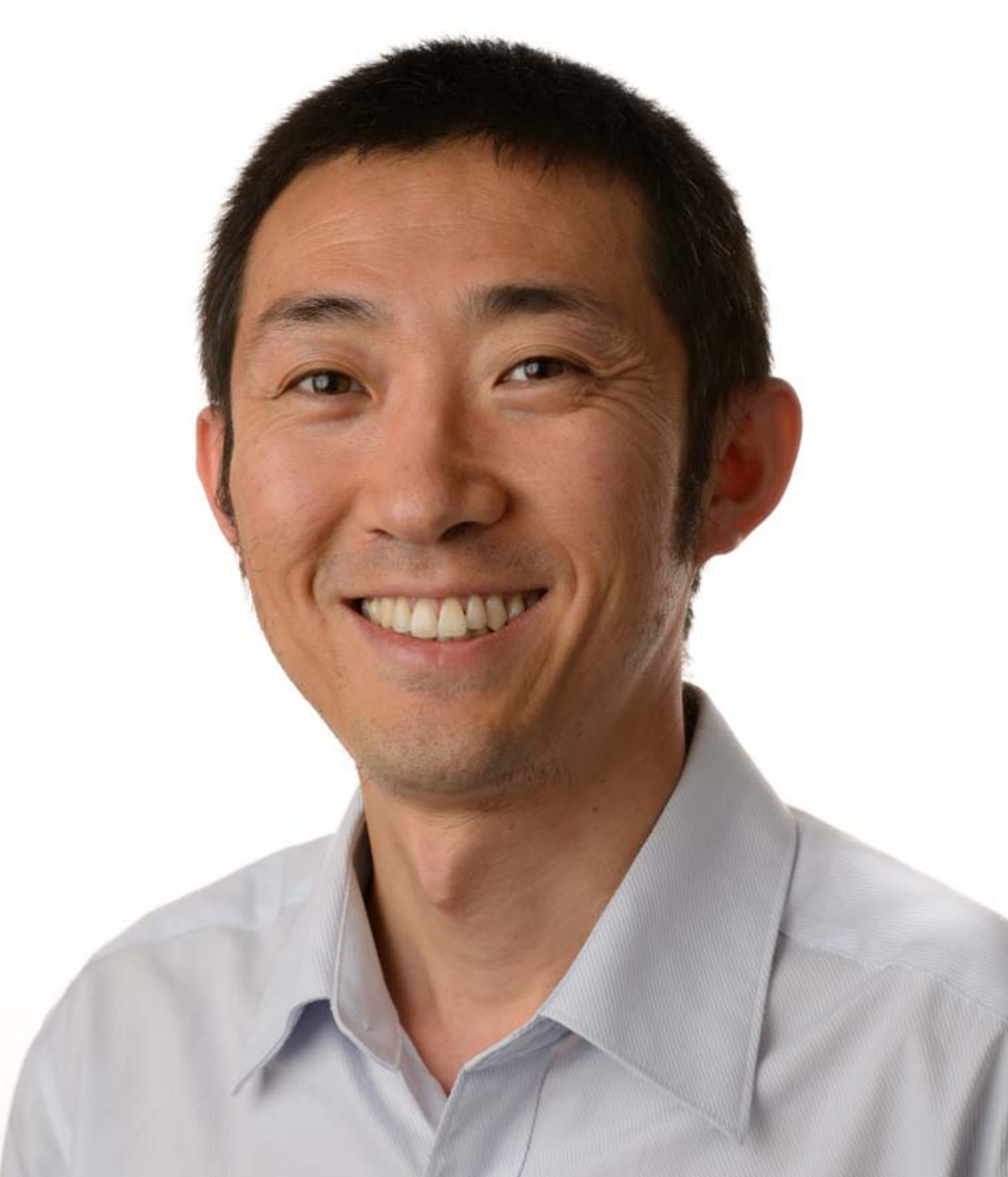}}]{Jin Ma} (M'06) received the B.S. and M.S. degree in Electrical Engineering from Zhejiang University, Hangzhou, China, the Ph.D. degree in Electrical Engineering from Tsinghua University, Beijing, China, in 1997, 2000, and 2004, respectively. From 2004 to 2013, he was a Faculty member of North China Electric Power University. Since September, 2013, he has been with the School of Electrical and Information Engineering, University of Sydney. His major research interests are load modeling, nonlinear control system, dynamic power system, and power system economics. He is the member of CIGRE W.G. C4.605 “Modeling and aggregation of loads in flexible power networks” and the corresponding member of CIGRE Joint Workgroup C4-C6/CIRED “Modeling and dynamic performance of inverter based generation in power system transmission and distribution studies”. He is a registered Chartered Engineer in UK.
\end{IEEEbiography}
\end{document}